\documentclass[10pt,a4paper,conference]{IEEEtran}

\usepackage{amsfonts,amsmath}

\usepackage{xspace}

\usepackage{url}
\usepackage{booktabs} 
\usepackage{xcolor}

\usepackage{mdframed}

\definecolor{vfl}{RGB}{0, 88,150}
\definecolor{ferrarired}{rgb}{1.0, 0.11, 0.0}
\definecolor{bluegreen}{rgb}{0.0, 0.5, 0.5}

\usepackage{graphicx}
\DeclareGraphicsExtensions{.pdf}

\usepackage{tabu}
\usepackage{paralist}
\usepackage{units}
\usepackage[np]{numprint}
\npstyleenglish

\newcommand{\figwidth}{1\columnwidth}
\usepackage{xcolor}

\newenvironment{mydef}%
  {\begin{mdframed}[backgroundcolor=black!5]}%
  {\end{mdframed}}

\newcommand{\E}[1][]{\ensuremath{{\rm{E}}\!\left[#1\right]}}

\newcommand{\Std}[1][]{\ensuremath{{\rm{Std}}\!\left[#1\right]}}

\newcommand{\Qx}[0]{\ensuremath{Q|_x}\xspace}

\def\BibTeX{{\rm B\kern-.05em{\sc i\kern-.025em b}\kern-.08em
    T\kern-.1667em\lower.7ex\hbox{E}\kern-.125emX}}
    
\usepackage{enumitem}    

\usepackage{balance}

\makeatletter
\IEEEtriggercmd{\reset@font\normalfont\normalsize\selectfont}
\makeatother
\IEEEtriggeratref{1}

\usepackage{tikz}
\usetikzlibrary{positioning}
\usetikzlibrary{arrows}
\usetikzlibrary{shapes.geometric}
\usetikzlibrary{arrows.meta,arrows}

\IEEEoverridecommandlockouts

\newcommand\copyrighttext{%
  \footnotesize \textcopyright 2020 IEEE. Personal use of this material is permitted.
  Permission from IEEE must be obtained for all other uses, in any current or future 
  media, including reprinting/republishing this material for advertising or promotional 
  purposes, creating new collective works, for resale or redistribution to servers or 
  lists, or reuse of any copyrighted component of this work in other works. This paper has been accepted for publication in the 4th International Workshop on Quality of Experience Management (QoE Management 2020), featured by IEEE Conference on Network Softwarization (IEEE NetSoft 2020).
	}
\newcommand\copyrightnotice{%
\begin{tikzpicture}[remember picture,overlay]
\node[anchor=south,yshift=10pt] at (current page.south) {\fbox{\parbox{\dimexpr\textwidth-\fboxsep-\fboxrule\relax}{\copyrighttext}}};
\end{tikzpicture}%
}

\begin{document}

\title{From QoS Distributions to QoE Distributions: a System's Perspective}

\author{
 \IEEEauthorblockN{Tobias Ho{\ss}feld$^1$, Poul E. Heegaard$^2$, Mart{\'\i}n Varela$^{3}$, Lea Skorin-Kapov$^4$, Markus Fiedler$^{5}$}
 \IEEEauthorblockA{$^1$Chair of Communication Networks, University of W\"urzburg, Germany\\
 $^2$NTNU, Norwegian University of Science and Technology, Trondheim, Norway\\
 $^3$Independent Researcher, Oulu, Finland \\
 $^4$Faculty of Electrical Engineering and Computing, University of Zagreb, Croatia\\ 
 $^5$Blekinge Institute of Technology, Dept. of Technology and Aesthetics, Karlshamn, Sweden
}
Email: $^1$tobias.hossfeld@uni-wuerzburg.de, $^2$poul.heegaard@ntnu.no,
\\
$^3$martin@varela.fi, $^4$lea.skorin-kapov@fer.hr, $^5$markus.fiedler@bth.se
\thanks{A script implementing the Beta distribution approximation and QoE metrics for arbitrary QoE measurements is published at Github:
https://github.com/hossfeld/approx-qoe-distribution}
}


\maketitle

\copyrightnotice
\pagestyle{plain}

\begin{abstract}
In the context of QoE management, network and service providers commonly rely on models that map system QoS conditions (e.g., system response time, paket loss, etc.) to estimated end user QoE values. Observable QoS conditions in the system may be assumed to follow a certain distribution, meaning that different end users will experience different conditions. On the other hand, drawing from the results of subjective user studies, we know that user diversity leads to distributions of user scores for any given test conditions (in this case referring to the QoS parameters of interest). Our previous studies have shown that to correctly derive various QoE metrics (e.g., Mean Opinion Score (MOS), quantiles, probability of users rating ``good or better'', etc.) in a system under given conditions, there is a need to consider rating distributions obtained from user studies, which are often times not available. In this paper we extend these findings to show how to approximate user rating distributions given a QoS-to-MOS mapping function and second order statistics. Such a user rating distribution may then be combined with a QoS distribution observed in a system to finally derive corresponding distributions of QoE scores. We provide two examples to illustrate this process: 1) analytical results using a Web QoE model relating waiting times to QoE, and 2) numerical results using measurements relating packet losses to video stall pattern, which are in turn mapped to QoE estimates.

\end{abstract}



\section{Introduction and background}\label{sec:modeling}

QoE management mechanisms deployed by system and network providers rely on models that map observable QoS performance measures to application level metrics, which may in turn be used to estimate overall Quality of Experience (QoE) as perceived by end users. Such models are derived from subjective user studies, where end users are exposed to various test conditions, such as different system QoS conditions (e.g., loss rates, delays), or different application level conditions (e.g., video stalls, web page load time) and asked to provide subjective rating scores. Ratings are commonly averaged to obtain a Mean Opinion Score (MOS), thus resulting in so-called \textit{QoS-to-MOS} mapping functions.   

In the context of subjective user studies, with different users perceiving both quality and value differently, \cite{zeithaml1988consumer}, \textit{user diversity} will inherently impact the distribution of rating scores for a given test condition \cite{janowski2009modeling,hossfeld2016qoe}. While a MOS value for a given test condition represents an ``average user'' rating, the drawback lies in averaging out user diversity, thus providing no insights into actual QoE distributions across users. 

From a QoE management point of view, previous work has argued that there is a clear interest among network/service providers in estimating the \textbf{distribution of QoE ratings in their system}, rather than just estimating a single MOS value \cite{hossfeld2016qoe}. Such insights into QoE distributions allow providers to derive metrics such as \textit{Good or Better} (GoB) ratio (giving the probability that for a given condition, the user rating will be ``good or better'' \cite{moller2000models}, corresponding to ratings 4 or 5 on a 5pt. Absolute Category Rating, ACR, scale), \textit{Poor or Worse} (PoW) ratio (the probability that user ratings will be 1 or 2 on a 5pt. ACR scale), or various quantiles. Depending on a provider's QoE management goals, such metrics can provide valuable input for QoE control and optimization, such as invoking resource (re)allocation mechanisms. 

In our previous work \cite{hossfeld2019fundamental}, we have tackled the challenge of how to derive a QoE distribution in a system, given: 1) the distribution of a system performance condition, and 2) the user rating distribution for fixed values of the condition as observed in subjective studies. We briefly summarize our previous findings as follows:

\begin{itemize}
    \item We proved a \textbf{fundamental relationship} showing that the \textbf{expected QoE} in the system can be derived by applying the MOS mapping function $f(x)$ to the measured QoS distribution $X$ in the system: $E[Q] = E[f(X)]$.
    \item We showed that to derive additional QoE metrics in the system it is \textbf{necessary to use corresponding mapping functions} derived from user rating distributions in subjective studies. E.g., to derive the GoB in the system, a QoS-to-GoB mapping function $g(X)$ is needed: $GoB[Q] = E[g(X)]$.
    \item We showed that to derive the complete QoE distribution in a system with a given system performance distribution, we need to know the distributions of rating scores observed in the subjective study per tested performance condition.
\end{itemize}

In realistic cases, studies for the most part report only QoS-to-MOS mapping functions, thus leading to the challenge of how to derive the aforementioned metrics. In this paper, we build on our previously reported results by proposing a \textbf{methodology for approximating user rating distributions}, given only a QoS-to-MOS mapping function and second order statistics. {We select the parametric \textbf{Beta distribution} and derive its parameters based on the MOS and standard deviation of opinion scores (SOS)}. We then illustrate how to combine this rating distribution with a system parameter distribution so as to obtain the QoE distribution in the system. 

The paper is organized as follows. 
Section~\ref{sec:methodology} illustrates the process of deriving QoE in a system, applying the previously mentioned fundamental relationship between QoE in the system and subjective user studies for arbitrary QoE metrics. Section~\ref{sec:web} provides analytical results using a Web QoE model relating waiting times to QoE, while Section~\ref{sec:video} gives  numerical results based on measurements relating packet losses to video stall frequencies and durations, which are in turn mapped to QoE estimates. Conclusions and outlook are given in Section~\ref{sec:conclusions}.

\newcommand{\sosa}{\theta}

\begin{figure*}[htpb]%
\centering
\tikzstyle{b} = [rectangle, draw, fill=blue!5, node distance=4.3cm, text width=7.2cm, align=left, rounded corners, minimum height=2.0cm, thick, inner sep=0.25cm]
\tikzstyle{innerb} = [rectangle, draw, fill=blue!5, node distance=4.3cm, text width=3.5cm, align=left, rounded corners, minimum height=2.0cm, thick, inner sep=0.25cm]
\tikzstyle{l} = [draw, -latex',thick]
\begin{tikzpicture}[auto]
    \node [b, fill=blue!10, text width=8.5cm] (system) {\textbf{System and services }
												\begin{itemize}
													\item utilization, request patterns, 
													\item configuration,
													\item implementation, $\dots$
												\end{itemize}
												};		
    \node [b, fill=blue!10, below of=system, node distance=3.5cm, text width=8.5cm] (R) {\textbf{System parameter distribution $X$} \\
									Parameter $X$ of the system is measured and is a (multi-dimensional) random variable (RV). \\
									$H(x) = P(X \leq x), h(x) = \frac{d}{dx} H(x)$												
																			};        
	
		\node [b,  fill=yellow!15, right of=system, node distance=9.3cm, text width=8.5cm] (subjective) {\textbf{Subjective studies}\\
													For fixed conditions $x$, user ratings (random variable, RV) are measured in a subjective study incl.
													 diversity of test subjects. 
												};
		\node [innerb, fill=yellow!15,  below of=subjective, node distance=3.5cm, text width=8.5cm] (Qt) {\textbf{Conditional rating distribution $\Qx$} 
		\begin{itemize}
		    \item empirical distribution 
		    \item parametric distribution (Sec.~\ref{sec:web}) 
		    \item Beta distribution approx. with $f(x)$ and $\theta$ (Sec.~\ref{sec:video}) 
		\end{itemize}};
    \path (R) -- node [below=2cm] (combine) {$\bigodot$} (Qt);   
		\node [node distance=1.5em, above of=combine] (textComb) {combine distributions};

    \path [l] ([xshift=-.5cm]system.south) --  node [right, text width=3.6cm, align=left] {random variable $X$, e.g. response time} ([xshift=-.5cm]R.north);
	\path [l] ([xshift=.5cm]subjective.south) --  node [left, text width=4.6cm, align=right]{e.g. MOS mapping $f(x)$\\ SOS parameter $\theta$} ([xshift=.5cm]Qt.north);

		\node [b, fill=green!15, below of=combine, node distance=2.0cm, text width=17.5cm] (systemQoE) 
							{\vspace{-1em}\center{\textbf{\large QoE distribution $Q$ in the system}}\\[0.5em] 
							The QoE distribution $Q$ over all users is 
							$Q(i) = P(Q\leq i) = \int_{x=0}^\infty Q(i|x) \cdot h(x) dx$. The distribution 
							$Q$ allows to derive metrics like expected QoE in the system, $E[Q]$, or GoB in the system $GoB[Q]$. 
							};	
	
		\path [l] (R) |- (combine);
	  \path [l] (Qt) |- node [right, pos=0.35, text width=4cm] {} (combine) ;
    \path [l] (combine) -- (systemQoE);
    
\end{tikzpicture} 
%
\caption{Overview on system QoE (being observed in a real system) and user rating distributions in a subjective study.}%
\label{fig:systemQoE}%
\end{figure*}
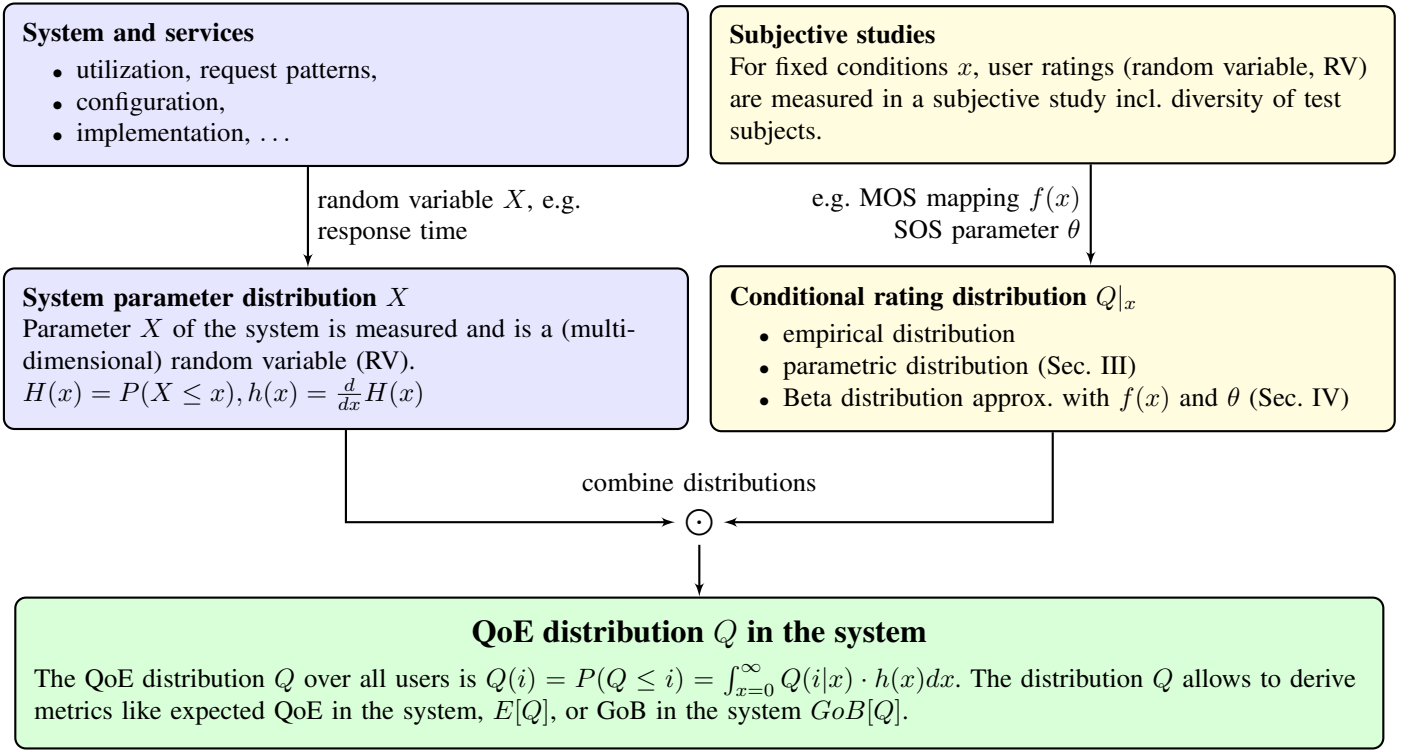

\section{Methodology} \label{sec:methodology}
Figure~\ref{fig:systemQoE} provides an overall picture of deriving QoE at the system level. In a system, its users will experience changes in system performance, caused by network effects (e.g., losses, response times, throughput), or application / service configuration (e.g., blurriness, number of stalls, stall duration, infrastructure capacity).
The system's performance depends on various \textit{system conditions}, including its configuration, its implementation, the system operational state, and the system utilization. 
Since the system utilization varies as the offered load varies, 
the users will experience different performance, in this paper represented by a (potentially multi-dimensional) random variable, which we write in scalar form $X$.
The cumulative distribution function (CDF), $H(x)$, and the probability density function (PDF), $h(x)$, of $X$ are:
\begin{equation}
H(x)=P(X\leq x), \; h(x) = \frac{d}{dx} H(x)
\label{eq:responseTime}
\end{equation}

Two different users experiencing the same (multi-dimensional) system condition (e.g., stall frequency and duration), $x$, may rate the situation differently due a variety of reasons (e.g., user mood, expectations, past experiences), which we refer to as \emph{user diversity}. 
This is represented by a random variable \Qx for the QoE user ratings, given the same system condition $x$, with the CDF $Q(i | x)$. 
\begin{align}
Q(i | x) &= P(Q\leq i | X=x)
\label{eq:qt}
\end{align}
We obtain a probability mass function (PMF) and a probability density function (PDF), $q(i | x)$, for discrete and continuous rating scales, respectively. 

The $H(x)$ might change due to reconfiguration or reimplementation of the system and its service, or due to changes in the offered load or system utilisation.
In user studies, the $Q(i|x)$ is typically obtained under certain (controlled or observed) system performance conditions, which do not reflect the $H(x)$ (i.e.,
current system performance distribution).

\subsection{Derivation of System QoE Distribution}
$Q$ is the 
random variable for the QoE user ratings over all the system performance conditions, with the CDF $Q(i)$
\begin{align}
Q(i) &= P(Q\leq i) = \int_x Q(i | x) h(x) dx 
\label{eq:entireDist}
\end{align}
which is obtained by integrating over the (potentially multi-dimensional) QoS parameter $X$. For example, if $X$ is two-dimensional, a double integral is to be computed.
We may directly derive the PMF and PDF for a discrete and continuous user rating scale, respectively. 
\begin{align}
q(i) = \int_{x} q(i | x) h(x) dx
\end{align}

Please note, when $X$ is a discrete distribution (as typically obtained in measurements, e.g., number of stalls, total stall duration discretized in bins), the sum is computed over all possible conditions of $x$
\begin{align}
Q(i) &= P(Q\leq i) = \sum_{x} Q(i | x) h(x) dx 
\end{align}
using the discrete PMF $h(x)$, see Section~\ref{sec:video} for an example of multidimensional discrete empirical measurements.

Knowing the distribution $Q$ allows us to derive various metrics of interest, such as the expected system QoE $\E[Q]$, the ratio of users rating Good-or-Better $GoB[Q]=P(Q\geq k)$ where $k$ indicates 'good' on the rating scale, or the ratio of users rating Poor-or-Worse $PoW[Q]=P(Q\leq j)$ where $j$ indicates 'poor' on the rating scale .

\subsection{Using Empirical Distributions from Subjective Studies}
The probabilities $Q(i | x)$ may be estimated from user ratings obtained by means of subjective studies, e.g., in the laboratory, via crowdsourcing, or by field trials, as long as the system condition $x$ is observed. 
If the set of $k$ ratings, $\mathbf{I}_x = \{ i_1, \cdots , i_k\}$, is available from a study under test condition $x$, then we can estimate the empirical distribution of $P(Q=i|X=x)$ from the data set,  
$\mathbf{I}_x$.

Besides empirical distributions, parametric models of the QoE distribution may be utilized. For example, literature has shown that for web QoE, the conditional user rating distribution $\Qx$ for a particular waiting time $x$ may be approximated with a binomial distribution \cite{hossfeld2016qoe}. This example of a parametric distribution will be considered in Section~\ref{sec:web}.


\subsection{Beta Distribution Approximation for $\Qx$}
Typically, in the literature, such a set or parametric distribution is not provided, but only the QoS-to-MOS mapping and the description of the test conditions used in the study.  We then have to use the QoS-to-MOS mapping, and second order statistics (standard deviation or sample variance) or the SOS-hypothesis \cite{hossfeld2011sos} (postulating a square relationship between MOS and standard deviation of opinion scores, SOS), and select an appropriate probability density function that fits the available statistics from the study. 

The main contribution of this paper is the answer to the following fundamental question:
\begin{mydef}
\textbf{Fundamental question:}
How can we derive QoE distributions in systems when observing QoS only? How can we approximate the QoE distribution in practice using existing QoS-MOS mapping functions? 
 \end{mydef}

We approximate the conditional QoE distribution $\Qx$ for condition $x$ by using a continuous Beta distribution $Z_0$ with parameters $a$ and $b$ which is defined on the interval $[0;1]$. For obtaining a QoE distribution in the range of $[L;H]$, e.g. the common 5-point scale $[1;5]$, the Beta distribution is linearly transformed.
\begin{equation}
    Z = (H-L)Z_0 + L \text{ with } Z_0 \sim Beta(a,b)
\end{equation}
Accordingly, the expected value is $\E[Z]=(H-L) \E[Z_0] + L $ and the variance is $Var[Z] = (H-L)^2 Var[Z_0]$.

The parameters $a,b$ of the the Beta distribution depend on the MOS value $m = \E[\Qx]$ and the corresponding standard deviation $s = \Std[\Qx]$ for condition $x$. While the MOS value $m=f(x)$ is obtained by the QoS-to-MOS mapping function, we derive the standard deviation by utilizing the SOS hypothesis \cite{hossfeld2011sos}. The SOS hypothesis provides a relationship between the mean opinion score (MOS) $m$ and the standard deviation of the opinion scores (SOS) $s$. Thereby, the SOS parameter $0 < \sosa < 1$ is independent of the particular condition and holds for a certain service, e.g. $\sosa=0.25$ is measured in subjective studies for web QoE \cite{hossfeld2016qoe}.
\begin{equation}
    s^2 = \sosa (H-m) (m-L)
\end{equation}

With the well known expected value and the variance of the Beta distribution $Z_0$, we are therefore able to derive the parameters $a$ and $b$ for a given MOS value $m$ and SOS parameter $\sosa$. It is $H>L$.
\begin{align}
    \E[Z_0] & = \frac{a}{a+b} = \frac{m-L}{H-L} \\
    Var[Z_0] &= \frac{ab}{(a+b)^2(a+b+1)} = \frac{\sosa(H-m)(m-L)}{(H-L)^2}
\end{align}
This is algebraically transformed and we obtain the parameters of the Beta distribution as a function of the MOS $m$ and given SOS parameter $\sosa$. 
\begin{align}
    a_\sosa(m) = \frac{(1-\sosa)(m-L)}{\sosa(H-L)} \quad
    & b_\sosa(m) = \frac{(1-\sosa)(H-m)}{\sosa(H-L)} \label{eq:beta:parameters}
\end{align}

As a final result of the Beta distribution approximation, we describe the conditional QoE distribution $\Qx$ for a MOS value $m=f(x)$ on a rating scale $[L;H]$ as Beta distribution with the parameters $a_\sosa(m)$ and $b_\sosa(m)$. Thereby, the SOS parameter $\sosa$ is given for the service under consideration and a MOS mapping function $f(x)$ is available.
\begin{align}
    Z_0 & \sim Beta(a,b) \text{ with } a=a_\sosa(f(x)), b=b_\sosa(f(x)) \\
    \Qx & = (H-L) Z_0 + L
\end{align}

The CDF of the Beta distribution is described with the incomplete Beta function $B(y,a,b)$ and the complete Beta function $B(a,b)$ with parameters $a,b$ as in Eq.(\ref{eq:beta:parameters}).
\begin{equation}
    \Qx(y) = P(\Qx \leq y) = \frac{B(\frac{y-L}{H-L},a,b)}{B(a,b)}
\end{equation}

Then we obtain the unconditional system QoE distribution $Q$ with the PDF $h(x)$ for the condition $X$ according to Eq.(\ref{eq:entireDist}). The CDF is defined for any $y \in [L;H]$.
\begin{align}
     Q(y) = P(Q \leq y) = \int_{x} \Qx(y) h(x) dx 
\end{align}


\section{Case study: Web QoE}\label{sec:web}
In the following, we consider a simple example to illustrate how to derive the QoE distribution in the system. 
We study a single web server offering users a certain service like access to a static site or authentication \cite{lorentzen2010authentication}. Literature provides web QoE models that may be utilized to map the waiting times to QoE. In particular, a recent web QoE model \cite{hossfeld2018:SpeedIndex} describes an exponential relationship between the speed index (SI) as a proxy for perceived
page load times (PLT) and MOS values. 

\subsection{QoE Model: Binomial Distribution}
The exponential MOS mapping function follows the IQX
hypothesis \cite{fiedler2010generic} and reveals a sensitivity parameter $\beta \approx  0.25$, cf. \cite{hossfeld2018:SpeedIndex}. This MOS mapping function $f(x)$ maps the response time $x$ of the web service  to a MOS value on a 5-point absolute category rating scale ranging from 1 (bad) to 5 (excellent).
\begin{equation}
    f(x) = n e^{-\beta x}+1 \label{eq:mapping}
\end{equation}
with $n=4$ due to the used 5-point rating scale, and $\beta=0.25$. 

Furthermore, the analysis of a subjective study on web QoE \cite{hossfeld2016qoe} revealed that the user ratings $\Qx$ can be accurately approximated by a binomial distribution for given $x$. Hence, a discrete QoE rating scale is used.
Thus, for any response time $x$ of the system, we may approximate the distribution of \Qx with a Binominal distribution, $\Qx \sim Bino(n,p)+1$ with the expected user rating $\E[\Qx]=n p+1$ corresponding to the MOS value $f(x)$, which allows to derive the parameter $p$ of the binomial distribution.
\begin{equation}
\Qx \sim Binom(n,p)+1 \, \text{ with }  p = \tfrac{f(x)-1}{n} = e^{-\beta x}
\label{eq:bino}
\end{equation} 
The probability that the user rating is $\Qx=i$ is
\begin{equation}
P(\Qx=i) = \binom{n}{i-1} e^{- \beta x (i-1)} (1-e^{-\beta x})^{n-i+1}
\label{eq:binoFinal}
\end{equation}
for $i=1,2,3,4,5$ and $n=4$, $\beta=0.25$. 
For the sake of simplicity, we use the term `waiting time' in the following when referring to the speed index as proxy for the user perceived waiting time. 

\subsection{Waiting Time Model: Lognormal Distribution}
Different configurations of the web service result with various waiting times. Those waiting times are assumed to follow a log-normal distribution which is observed in practice for system response times, e.g. Enterprise Resource Planning (ERP) transaction systems \cite{mielke2006elements}. Please note that the actual distribution is not relevant for the illustration. The log-normal distribution has two parameters $\mu$ and $\sigma$ which may be derived for given mean $\E[X]$ and standard deviation $\Std[X]$ of the waiting time $X \sim LOGN(\mu, \sigma)$. The PDF of the log-normal distribution is
\begin{equation}
    h(x) = \frac{1}{x \sigma \sqrt{2 \pi}} exp\left( - \frac{(\ln x -\mu)^2}{2\sigma^2}\right) \label{eq:logn}
\end{equation}

Figure~\ref{fig:waiting:pdf} shows the PDF $h(x)$ for the three different system configurations with same mean waiting times $E[X]=\unit[4]{s}$, but different standard deviations (\unit[2]{s}, \unit[4]{s}, \unit[8]{s}).

\subsection{Deriving the QoE Distribution in the System}
\begin{figure}[t]
    \centering
    \includegraphics[width=\figwidth]{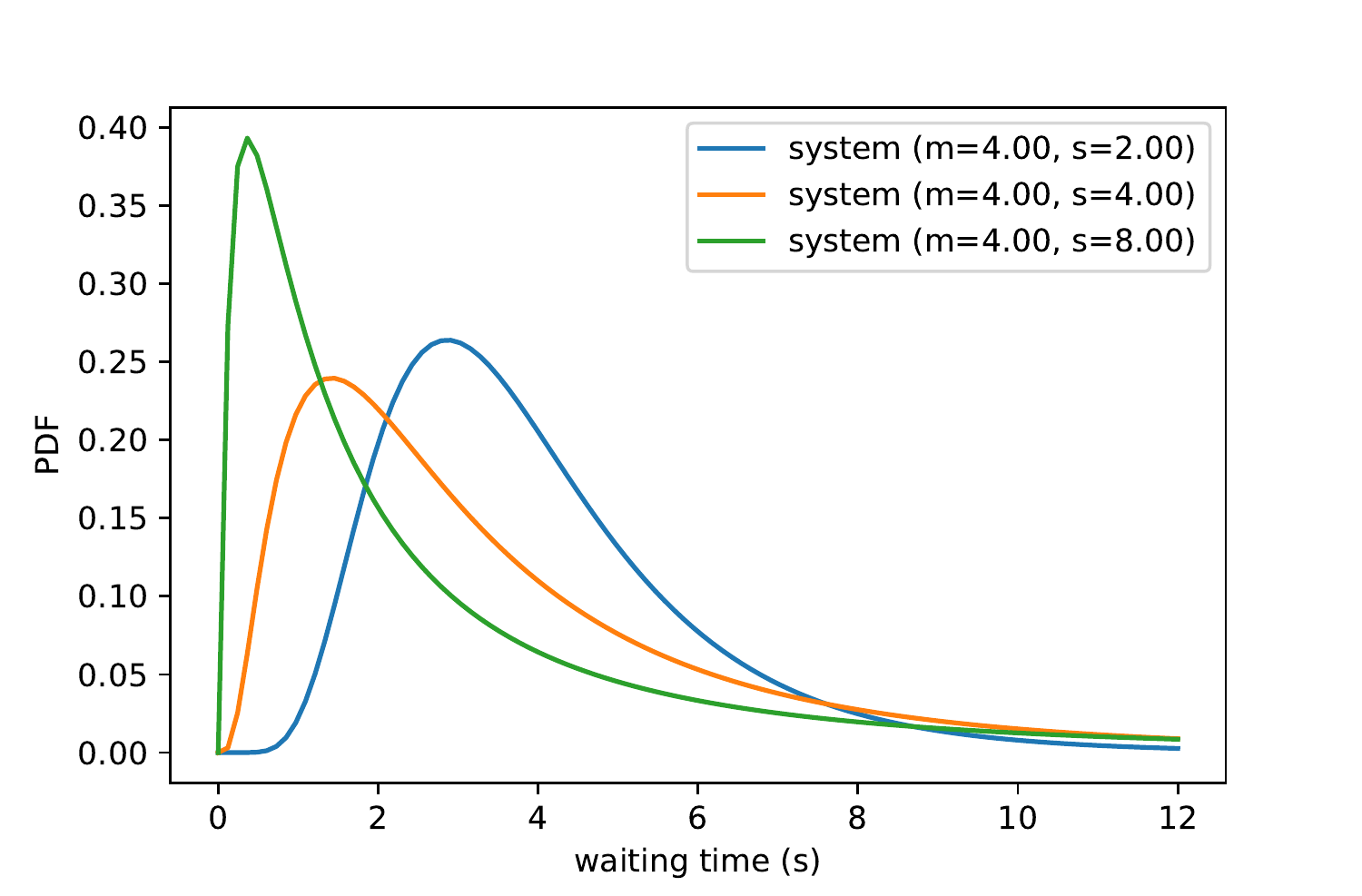}
    \caption{\emph{Web QoE example:} The three curves indicate different systems with same mean waiting time of $m=\unit[4]{s}$, but varying standard deviation $s$ which impacts the tail of the distribution. 
    }
    \label{fig:waiting:pdf}
    \centering
    \includegraphics[width=\figwidth]{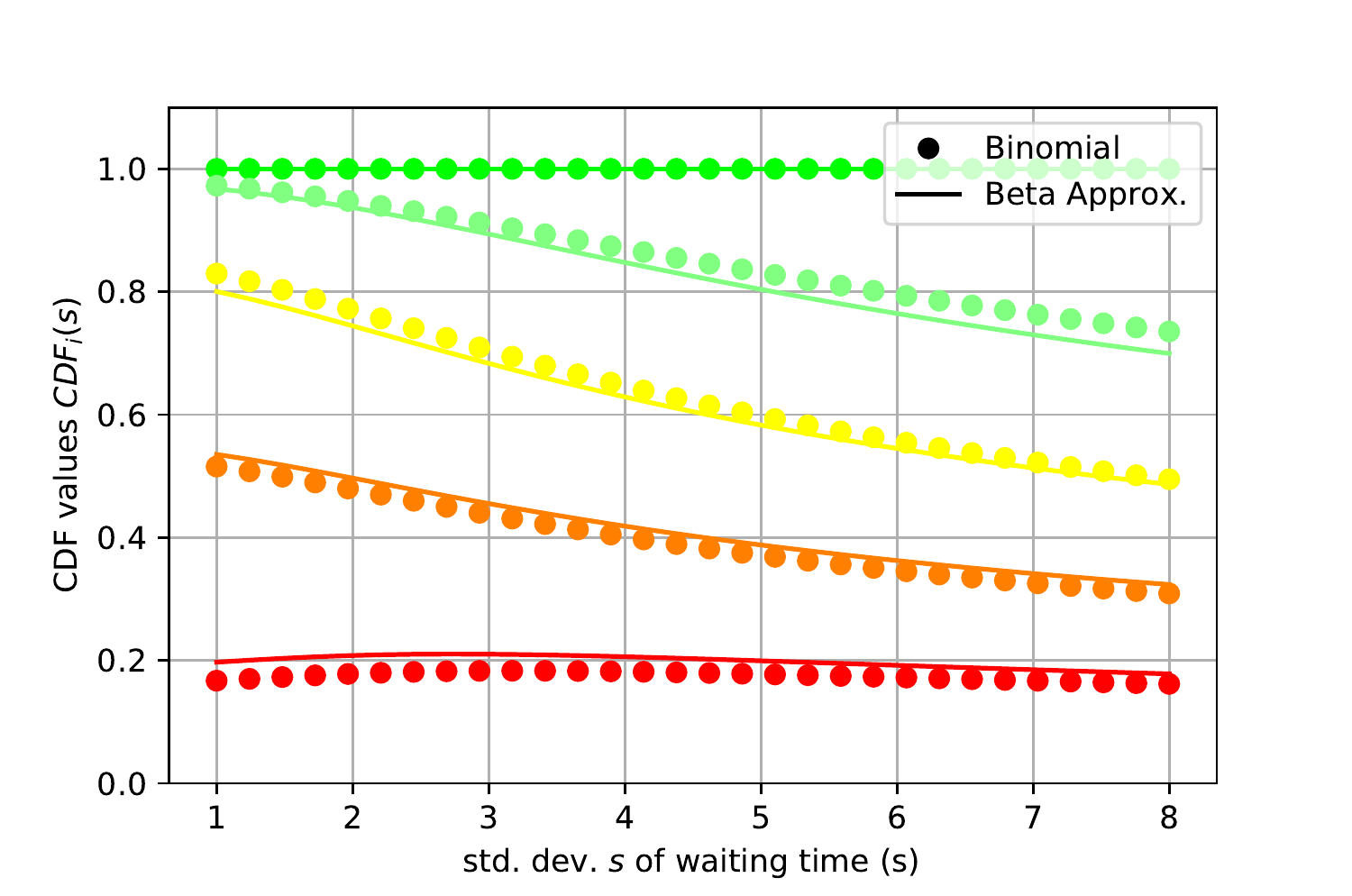}
    \caption{\emph{Web QoE example:} Approximation of system QoE distribution with Beta distribution. The different lines correspond to the CDF values $CDF_i(s)$ for $i=1$ (bad, red), 2 (poor, orange), 3 (fair, yellow), 4 (good, light green), 5 (excellent, green). The expected waiting time is $\E[X]=\unit[4]{s}$, while its standard deviation $s$ is varied on the x-axis. Thus, the lines correspond to $CDF_i(s) = P(\widehat{Q}_{m,s}\leq i)$ where the discrete QoE distribution $\widehat{Q}_{ms}$ is observed for a system with $E[X]=m, Std[X]=s$.
    }
    \label{fig:waiting:approx}
\end{figure}

Based on the fundamental relation (Eq.(\ref{eq:entireDist})) in Section~\ref{sec:methodology}, the QoE distribution in the system can be derived as follows:
\begin{align}
P(Q=i)   & = \int_{x=0}^\infty P(\Qx=i) h(x) dx \; , \, \forall i=1,\dots,5 
\end{align}		
This integral is derived numerically using Eq.(\ref{eq:binoFinal}) and Eq.(\ref{eq:logn}). 
%
Figure~\ref{fig:waiting:approx} shows the QoE distribution as filled circles, depending on the standard deviation of the waiting time. From the QoE distribution, we can easily observe the ratio of users in the system with poor or bad QoE (PoW ratio) as well as good or better QoE (GoB ratio). Interestingly, the GoB ratio increases with increasing variance in the system. This seems to be counter-intuitive, but log-normal distributions with the same expectations and growing variance get more heavy-tailed, cf. Figure~\ref{fig:waiting:pdf}, which usually indicates worse performance. However, a higher variance implies even more users with shorter waiting times, which improve the GoB ratio. On the other hand, a couple of users experience very long waiting times, which do not imply big differences in (anyway bad) QoE according to the exponential mapping, but yet a sinking PoW ratio as the variance of the waiting time increases.

\subsection{Beta Distribution Approximation}
Figure~\ref{fig:waiting:approx} compares the CDF values of the discrete QoE distribution $Q$, derived from the Binomial distribution, with the CDF values of the continuous Beta distribution $Q^*$, which is discretized through rounding and bounding to the scale $[1;5]$,
\begin{equation}
    \widehat{Q} = [Q^*] 
\end{equation}
with probability mass function 
\begin{align}
    \widehat{Q}(i) &= P(\widehat{Q}=i) = P(i-0.5 < Q^* \leq i+0.5) \\ \nonumber
    &= Q^*(i+0.5) - Q^*(i-0.5)
\end{align}
for $i=1,\dots,5$.
In Figure~\ref{fig:waiting:approx}, the exact CDF values of the QoE in the system are depicted as filled circles. The results from the Beta distribution approximation are plotted as solid lines with the SOS parameter $\theta=\frac{1}{n}=0.25$, which is derived by solving $\sosa(5-np)(np-1)=np(1-p)$. It can be seen the that differences are negligible and that the Beta approximation is sufficient in practice to derive the QoE distribution.




\section{Case study: Video streaming QoE}\label{sec:video}
We illustrate how our approach might be applied in a more complex use case, by considering a layered approach, where the system's QoS performance is first mapped to probability distributions for application-layer KPIs, and then those are used to estimate the QoE distribution itself. The performance model is taken from~\cite{varela2017embracing}. In that paper, a model is provided for deriving distributions of the number of stalls and their duration for non-adaptive HTTP streaming. In spite of its limitation in terms of lack of adaptation, the approach can be applied to more complex scenarios, as it allows to decouple the models~\cite{DBLP:journals/pik/VarelaSGF14}: instead of a single QoS to QoE mapping, we now have a QoS to application-layer KPI mapping, and then another mapping from this level to QoE. 

\subsection{Measurement Setup}
The experiment in~\cite{varela2017embracing} involved a large number of parameters, including loss ratio, loss burstiness, available bandwidth, buffer size, and content (different movies). An emulated network was setup, and the videos were streamed through it to a Dynamic Adaptive Streaming over HTTP (DASH) play-out emulator \cite{maki2015layered}, which recorded stalling events and their durations. From those measurements, two neural networks were trained to estimate the distribution of the number of stalls, as well as the stall durations for each experimental condition. From these results, we obtain a joint two-dimensional application-level KPI (QoS) distribution $X$ as visualized in Figure~\ref{fig:video:qos}; this joint QoS distribution is mapped to MOS according to the QoE model below.

\subsection{Derivation of QoE Distribution}
The QoE model consists of two parts. First, the application-level QoS parameters are mapped to MOS values. Then, for a certain MOS value and condition $x$, the entire QoE distribution is derived based on the Beta distribution approximation with a SOS parameter $\sosa$. In \cite{hossfeld2011quantification}, an SOS parameter of $\sosa=0.3$ was observed for non-adaptive video streaming based on subjective crowdsourcing studies with arbitrary devices from end users, while \cite{maki2015layered} reported $\sosa=0.1$ in a controlled laboratory study. 

The MOS mapping function is provided in \cite{hossfeld2013internet} based on subjective studies of non-adaptive streaming 
and follows an exponential relationship. This model takes into account the number $n$ of stalls and the total stall duration $t$ normalized by the video duration $d$ (in seconds). In the experimental results, videos of $d=\unit[60]{s}$ are considered.
\begin{equation}
  f(n,t) = 3.5 e^{-4.5\cdot t/d-5.7 \cdot n/d}+1.5    
\end{equation}

For the condition $x=(n,t)$, the QoE user rating distribution is approximated for the MOS $m=f(n,t)$ with the Beta distribution and the parameters $a_\sosa(m)$ and $b_\sosa(m)$ as defined in Eq.(\ref{eq:beta:parameters}). The corresponding CDF is then given by $\Qx(y)$ for any $y \in [1;5]$.

Then, the CDF of the QoE distribution $Q$ in the system is numerically derived. Please note that the number of stalls is a discrete random variable. Furthermore, the total stall duration is discretized in time slots of $\unit[1]{s}$, also yielding a discrete random variable.
\begin{equation}
    Q(y) = \sum_{n=0}^\infty \sum_{t=0}^\infty Q|_{(n,t)}(y) f(n,t) \, , \, y \in [1;5]
\end{equation}

\subsection{Results: Impact of loss rate}

\begin{figure}
    \centering
    \includegraphics[width=\figwidth]{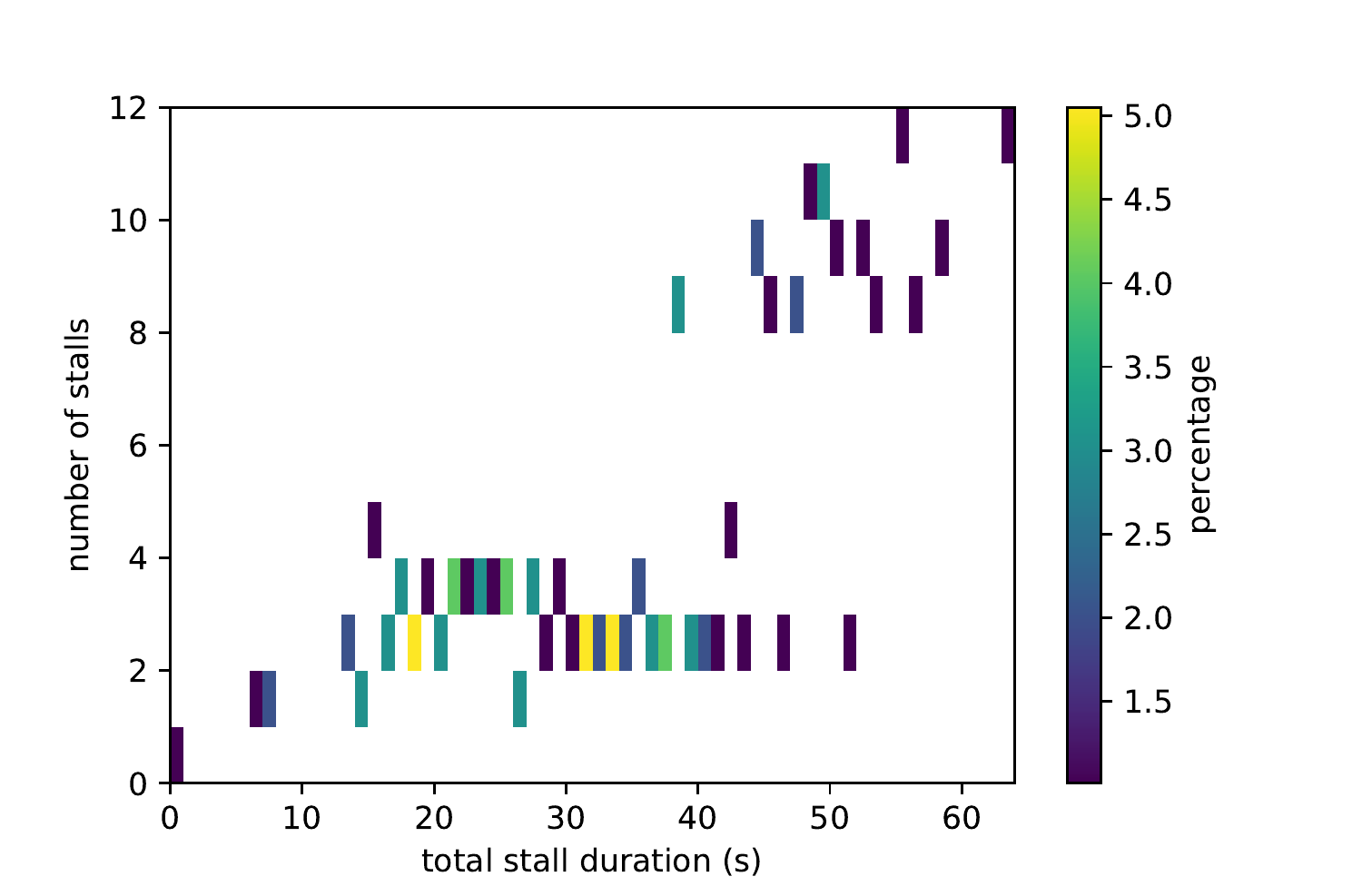}
    \caption{\emph{Video example:} Empirical joint probability mass function $h(n,t)$ of the QoS conditions (i.e. number $n$ of stalls and total stall duration $t$) for a scenario with a packet loss ratio of 9.6\%. The color indicates the joint probability $h(n,t)$.}
    \label{fig:video:qos}
    \centering
    \includegraphics[width=\figwidth]{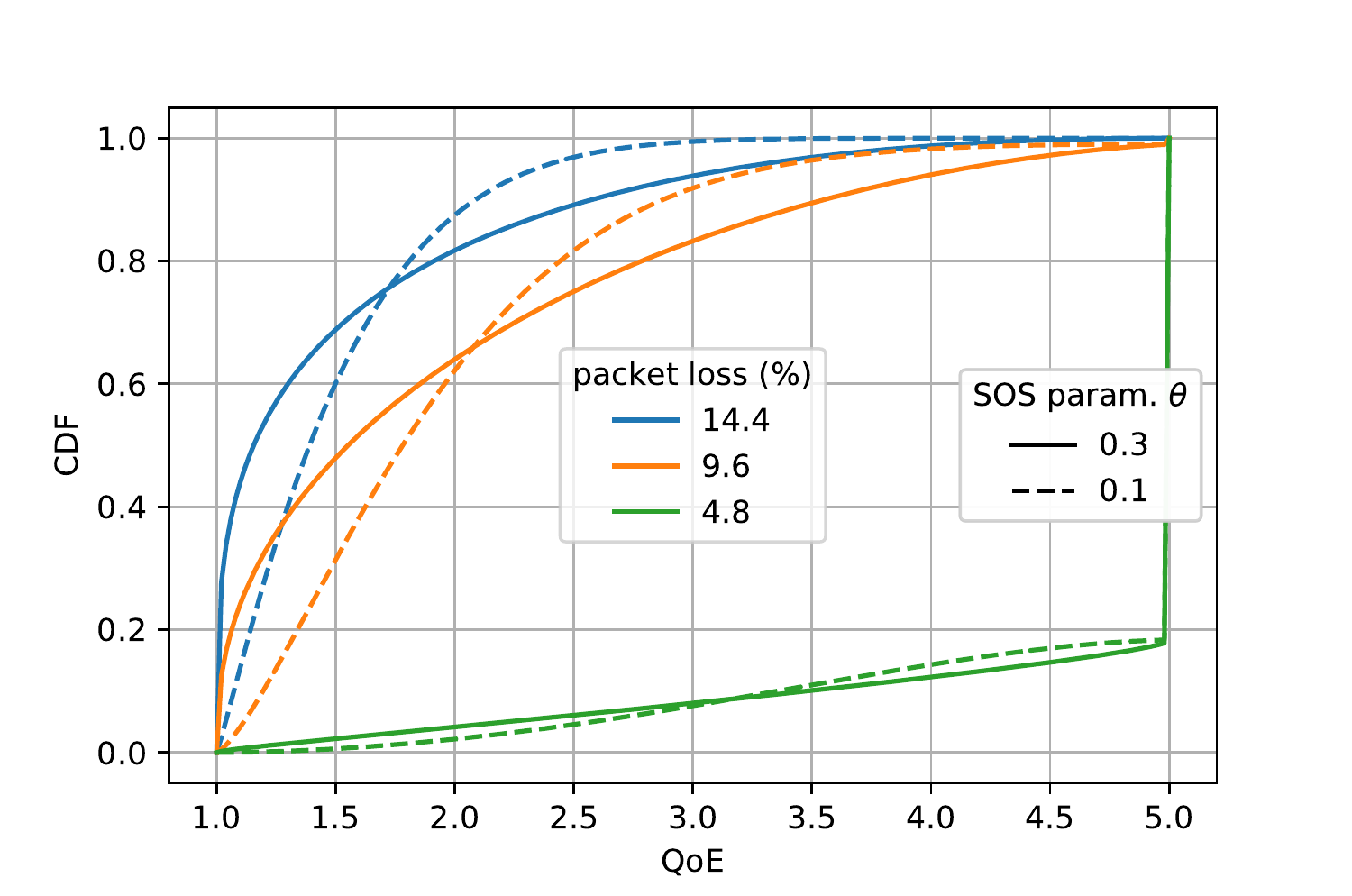}
    \caption{\emph{Video example:} QoE distribution depending on the packet loss ratio and the SOS parameter reflecting the user diversity.}
    \label{fig:video:qoe:sosa}
\end{figure}

Figure~\ref{fig:video:qoe:sosa} plots the CDF of the QoE distribution for different packet loss ratios (4.8\%, 9.6\%, 14.4\%). The solid lines correspond to a high user diversity with a SOS parameter of $\theta=0.3$, while the dashed lines are related to $\theta=0.1$, which are inline with the crowdsourcing \cite{hossfeld2011quantification} and laboratory results \cite{maki2015layered}. First of all, we observe that there is no big difference of the curves for the different SOS parameter values reflecting the user diversity, see also the GoB and PoW ratios in Table~\ref{tab:video:qoe}. Hence, if the SOS parameter is not known in practice, it may be sufficient to derive the numerical results for both values $\sosa=0.3$ and $\sosa=0.1$.

Figure~\ref{fig:video:qoe:sosa} reveals significant changes in the shapes of the QoE CDFs as the packet loss rate grows. While for a packet loss of 4.8\,\%, GoB dominates (due to a majority of top ratings of about 82\%) and PoW is in the order of some few percent, the CDFs become concave as loss increases, with significant PoW above 60--80\,\%, and GoB in the order of some few percent. The QoE distributions clearly indicate a change of the operational regime from a (rather) good state (packet loss of 4.8\,\%) to a bad state (higher packet loss of 9.6\,\% and above).

\begin{table}[tb]
    \centering
    \caption{Video example: GoB $P(Q\geq 4)$ and PoW $P(Q\leq 2)$ for different packet loss ratios and SOS parameters $\sosa$.}\label{tab:video:qoe}
    \begin{tabular}{lllll}
        \toprule
        loss & GoB & GoB  & PoW & PoW \\
        (\%) & $\sosa$=0.3 & $\sosa$=0.1 & $\sosa$=0.3 & $\sosa$=0.1 \\
         \midrule
4.8 & 0.88 & 0.86  & 0.04 & 0.02 \\
9.6 & 0.06 & 0.02  & 0.64 & 0.61 \\
14.4 & 0.01 & 0.00  & 0.81 & 0.87 \\
        \bottomrule
    \end{tabular}
\end{table}

\section{Discussion and Conclusions} \label{sec:conclusions}
For service providers, QoE can be a much more complex issue than just obtaining
a nice, MOS-like indicator of how quality looks like for their service(s). For
many interesting applications, ranging from pricing to monitoring to resource
optimization, it is important to take user diversity into account. Ideally,
service providers would have access to the distribution of QoE across their
service's users. This is unfortunately not possible in most cases, as most
modeling efforts focus on QoS-to-MOS type mappings.

In previous work, we showed that there exists a fundamental relationship between
the QoE in the system and the distribution of MOS values (in that the expected
QoE in the system will be the same as the expected MOS). We further showed that
while deriving MOS distributions is possible, these do not usually correspond to
the actual distribution of QoE in the system, and that if other, more
interesting metrics than MOS (e.g., GoB ratio) are to be used, then the modeling
efforts should be extended to provide QoS-to-M type mappings, with M being the
desired metrics.

In this paper we have gone one step beyond, and propose a method for estimating
rating distributions (i.e., QoE distributions), provided we have a QoS-to-MOS
mapping and the SOS parameter for that service. We used the Beta distribution,
and showed how to derive its parameters from the MOS and SOS. We then showed how
to combine this rating distribution with that of the system's QoS parameters, to
finally derive an estimate of the ratings distribution in the system.

We have illustrated the approach with two case studies. The web QoE example 
provides a ``mathematically nice'' analytical
example, in which the distribution of QoS parameters is assumed to be known. The
HTTP video streaming example 
shows how this can this be achieved in cases
where the relevant QoS parameters' distribution is not known in advance. 
The results in this paper provide a solution to the problem of understanding the
QoE distribution in a system, in cases where the necessary data is not directly
available in the form of models going beyond the MOS, or where the full details
of subjective experiments are not available.

\hfill

\section*{Reproducibility and Open Access}
A script implementing the Beta distribution approximation and QoE metrics for arbitrary QoE measurements is published at Github:
https://github.com/hossfeld/approx-qoe-distribution

\hfill
\section*{Acknowledgments}

This work was partly funded by NTNU hosting the Third International Symposium on Quality of Life (QoL-2020) in Trondheim, Norway, in February 2020.
L. Skorin-Kapov's work has been supported in part by the Croatian Science Foundation under the project IP-2019-04-9793 (Q-MERSIVE).


\bibliographystyle{IEEEtran}
{\normalsize
\balance
\bibliography{literature}
}


\end{document}